\documentclass[format=acmsmall, review=false, screen=true]{acmart}
\usepackage{booktabs} 

\usepackage{amsmath}
\allowdisplaybreaks
\usepackage{amssymb}
\usepackage{amsthm}

\newtheorem*{mt-theorem*}{Matrix-Tree Theorem}{\bfseries}{\itshape}
\newtheorem{theorem}{Theorem}

\usepackage[ruled]{algorithm2e} 
\SetKwInOut{Input}{input}\SetKwInOut{Output}{output}
\SetKwFor{RepTimes}{repeat}{times}{end}

\usepackage{hyperref}
\hypersetup{
	colorlinks=true, 
	linktoc=all,     
	linkcolor=black,  
	citecolor=black,
	pdfpagemode=UseOutlines,
	breaklinks=true
}

\usepackage{mathtools}
\DeclarePairedDelimiter\abs{\lvert}{\rvert}%
\makeatletter
\let\oldabs\abs
\def\abs{\@ifstar{\oldabs}{\oldabs*}}
\let\oldnorm\norm
\def\norm{\@ifstar{\oldnorm}{\oldnorm*}}
\makeatother
\DeclareMathAlphabet{\mathcal}{OMS}{cmsy}{m}{n}
\SetMathAlphabet{\mathcal}{bold}{OMS}{cmsy}{b}{n}
\newcommand{\bigO}{\mathcal{O}}

\SetAlFnt{\small}
\SetAlCapFnt{\small}
\SetAlCapNameFnt{\small}
\SetAlCapHSkip{0pt}
\IncMargin{-\parindent}

\setcopyright{none}
\renewcommand\footnotetextcopyrightpermission[1]{}

\usepackage{soul}
\usepackage{color}
\soulregister\hl7
\soulregister\cite7
\soulregister\ref7

\begin{document}
\title{Random Spanning Trees for Expanders, Sparsifiers, and Virtual Network Security}
\subtitle{(Preliminary Version)}

\author{Shlomi Dolev}

\affiliation{%
  \institution{Ben-Gurion University of the Negev}
  \streetaddress{P.O.B. 653}
  \city{Beer-Sheva}
  \postcode{8410501}
  \country{Israel}}
\email{dolev@cs.bgu.ac.il}

\author{Daniel Khankin}
\affiliation{%
	\institution{Ben-Gurion University of the Negev, Israel}
	\streetaddress{P.O.B. 653}
	\city{Beer-Sheva}
	\postcode{8410501}}

\email{danielkh@post.bgu.ac.il}

\begin{abstract}
This work describes probabilistic methods for utilizing random spanning trees generated via a random walk process. Goyal et al. showed that the union of random spanning trees approximates the expansion of every cut of a graph. First, we generalize the method by Goyal et al. for weighted graphs and show that it is possible to approximate the expansion of every cut in a weighted graph with the union of random spanning trees generated by a random walk on a weighted graph. Second, we show that our union of random spanning trees is a spectral sparsifier of the graph. Moreover, we show that $\bigO(\log n /\epsilon^2)$ random spanning trees are required in order to spectrally approximate a bounded degree graph. This result closes a previously open question on the number of random spanning trees required for saprsification. Third, we show that our random spanning trees based construction provides security features for virtual networks, in the context of Software-Defined Networking.
Network virtualization coupled with Software-Defined Networking allows new on-demand management capabilities. We demonstrate such a service, namely, on-demand efficient monitoring or anonymity. The proposed service is based on network virtualization of expanders or sparsifiers over the physical network. The defined virtual (or overlay) communication graphs coupled with a multi-hop extension of Valiant randomization based routing lets us monitor the entire traffic in the network, with a very few monitoring nodes. We propose methods that theoretically improve services provided by existing monitoring or anonymity networks, and optimize the degree of monitoring or anonymity.
%
%
%
\end{abstract}

%
%

%
%


\maketitle
\thispagestyle{empty}



\section{Introduction}
\label{sec:introduction}

Today, companies rely heavily on networking and Internet connectivity for even the simplest of tasks. As such, traffic monitoring is a common requirement in private corporate networks. Monitoring a network requires monitors that are located at the nodes of a network. Monitors are not required to be located at each node of the network but can be placed at selected nodes. The preferred location for a monitor is governed by the amount of traffic passing through that location and the importance of it to the network routing paths. Thus, using the least number of monitoring nodes and their efficient exposure to network traffic is of high interest to \textit{network monitorability}.

On the contrarily, \textit{anonymity} may be of importance to particular users of the network. The anonymity of a subject is defined as being not identifiable within the anonymity set~\cite{pfitzmann_anonymity_2001}. The anonymity set is the set of all possible subjects who might cause an action. A subject is identifiable if we can get a hold of information that can be linked to.

The reader will not be wrong in considering anonymity and monitoring as two \textit{reciprocal} services. Indeed, as we imply later in this work, anonymity and monitoring are two interchangeable services, i.e., a trade-off exists between monitorability and anonymity in the network. In the discussion that follows, the considerations and the following conclusions apply to both anonymity and monitorability.

In this work, we are interested in creating a network service that will be able to provide monitorability or anonymity \textit{network features}. Moreover, due to the advent of flexibility in networking and services (e.g., cloud-based services), we are interested in a dynamic service that will be able to construct and reconfigure the services in an \textit{on-demand} manner.

The network service construction should be decoupled from the underlying physical infrastructure and traditional network protocols. In particular, due to the decentralized nature of traditional networking, it will be highly complex to design and construct a network service that will be able to provide the discussed features. Moreover, we need means to modify and re-configure the network directly.

In recent years, a new and an emerging paradigm gains acclaim; Software-Defined Networking (SDN) promises to shift the existing limitations of traditional networks. SDN allows the decoupling of the control plane from the data plane. With this separation, the network switches become simple forwarding devices, while the control logic is implemented at a logically centralized controller~\cite{kreutz_software-defined_2015}. Forwarding decisions are now flow based instead of destination based. An important characteristic of SDN is that it allows the network to be programmable through software applications running on top of the controller. 
	
The logical centralization provided by SDN enables simpler, through high-level languages, modifications to network policies. Additionally, such control logic centralization in a controller, allows the latter to have global information of the network state. Those features of SDN highly simplify the development of sophisticated network services and applications.
	
SDN is a building block and mechanism for realizing network virtualization (NV) and architecture independent network functions virtualization (NFV). Network virtualization allows us to create virtual networks which network's topology is decoupled from the topology of the underlying physical network, and dynamically create policy-based virtual networks \cite{rao_sdn_2014}. An overlay network is one of many possible forms of network virtualization. Its main idea is to encapsulate a network service decoupled from the underlying infrastructure \cite{rao_sdn_2014}.

Network functions virtualization concerns implementation of network functions in software. Due to increasing demand for dynamic network architectures, network virtualization technology was utilized in Network-as-a-Service (NaaS) models that enabled dynamic deployment of a network service on-demand \cite{boubendir_-demand_2016}. Integration of SDN and NFV enables flexible, programmable, and dynamic deployment of network services \cite{boubendir_naas_2016}. 

Precisely speaking, we are interested in constructing an overlay network that can be deployed, re-configured, or shut down on-demand, and which is able to provide anonymity or monitoring networking as a service. We find that SDN paradigm is most suitable for the task.

One question, though, remains. Which network topology should we employ? Network topology can affect many networking factors, in particular, on the \textit{amount} of anonymity or monitorability in the network. For example, it has been shown in \cite{diaz_impact_2010} that topology of the network has an impact on the level of anonymity. One may consider that a complete graph based topology would best suit our needs. The motivation for this originates in the understanding that a message which arrives at a node could have originated from any node in the network (assuming the nodes in the path of a message are chosen at random) \cite{danezis_mix-networks_2003}. However, it was shown that having fewer routes results in better mixing of communication paths~\cite{diaz_impact_2010}, and therefore, in higher anonymity. 

Moreover, in order to conceal the relationships between incoming and outgoing flows, \textit{dummy traffic} must be added to the data flows. Dummy traffic is the coupling of \textit{dummy packets} to data packets leaving a node. The dummy packets are indistinguishable from the data packets (due to encryption of both). The complete graph based topology results in a considerable overhead when dummy traffic is used due to $\bigO(n)$ degree of the graph. Therefore, such topology will not be scalable~\cite{danezis_mix-networks_2003}. Diaz et al. have shown in \cite{diaz_impact_2010} that the overhead in using dummy traffic in a complete graph based topology is not only higher but possible to stay circulating in the network. For now, we can be certain that network topology in interest should have a restricted degree.

Network properties that provide randomization, such as random routing or mixing, contribute to higher anonymity \cite{danezis_mix-networks_2003, chaum_untraceable_1981, chaum_dining_1988, dolev_xor-trees_2000, hermoni_rendezvous_2014, valiant_scheme_1982}. However, in order to efficiently employ those and notably random routing, the topology should provide fast mixing property. Therefore, not any restricted topology that may provide high anonymity is suitable.

Further, both anonymity and monitoring services are highly dependent on the robustness of the network. Those services are \textit{path oriented}. As such, many existing attacks on those services are link-based attacks. In those attacks, the user is either forced to use a particular path or denied from using a path by causing a link failure on it. Consequently, in order to provide high reliability for those services, \textit{path diversity} and multipath routing should be supported as well.

The last thing to consider is the link capacity. For our needs, it is sufficient to define the capacity of a link as the amount of traffic that can flow through that link. Anonymity and monitoring services choose their locations and routing paths with a bias towards higher capacity \cite{dingledine_tor:_2004, dolev_routing_2010}.

To summarize, we are interested in a scalable overlay network, which provides path diversity, is robust, sparse, fast mixing, and exploits most of the capacity of the underlying network. Danezis has shown many benefits of exploiting expander graphs for anonymity networks~\cite{danezis_application_2009}. We consider that an overlay network, which is built on top of a physical network, should not only capture the expansion of the underlying network (assuming the underlying network is expander-like) but should also be sparser. 



Goyal et al.~showed in \cite{goyal_expanders_2009} (and later Frieze et al.~in \cite{frieze_expanders_2014}) that it is possible to construct an expander graph via random spanning trees. We follow this method of construction for our overlay network. However, we employ this method on weighted graphs in order to build a capacity biased overlay network. 


One method of random spanning tree construction is to use a random walk \cite{goyal_expanders_2009, frieze_expanders_2014, broder_generating_1989}. Given a graph $G=(V,E)$, a random walker starting from a random vertex travels the graph by selecting a random neighbor and moving there. Each time it arrives at an unvisited vertex it adds the edge through which it arrived at to the random spanning tree construction. The process continues until all vertices are visited. In the case of a weighted graph, the random walk is weighted such that its neighbor selection is proportional to the weights of connection edges. Such a method is easy to implement in SDN context. More than that, it is possible to distribute the overlay network construction among several SDN controllers. Each SDN controller will construct a different random spanning tree. 

The constructed random spanning trees are united in order to form an overlay network. The required number of random spanning trees is dependent on the desired expansion approximation, up to a factor of $\log n$.

Further, as our goal was to create sparse graphs, we show that the constructed overlay network graph approximates the underlying graph on the same set of vertices. Namely, the constructed graph spectrally sparsifies the underlying graph. The notion of spectral sparsification was introduced by Spielman and Teng in \cite{spielman_spectral_2011}, deriving a stronger notion than cut sparsifiers introduced by Bencz\'ur and Karger in \cite{benczur_approximating_1996}. They constructed sparsifiers with $\bigO(n \log ^c n)$ edges, where $c$ is a large constant. Later in \cite{spielman_graph_2011}, Spielman and Srivastava presented a nearly-linear time algorithm for graph sparsifiers with $\bigO(n \log n / \epsilon^2)$ edges. We show that our method of construction is equivalent to the sparsification method used in~\cite{spielman_graph_2011}. Though the method for construction in \cite{spielman_graph_2011} is faster, it may result in a disconnected graph \cite{batson_spectral_2013}. Besides, the method by Spielman and Srivastava requires solving $\bigO(\log n)$ linear equations, which computationally may not be possible or overwhelming for some (network) devices.  While our method of using a random walk is considerably computationally simpler.

Importantly, previous results on sparsification via random spanning trees left open the question of whether it is possible to spectrally approximate a graph with a few random spanning trees~\cite{batson_spectral_2013}. We answer this question and show that $\bigO(\log n /\epsilon^2)$ trees are sufficient in order to spectrally approximate a bounded degree weighted graph. 


We show that our network architecture is valuable for monitorability and anonymity. In order to introduce randomized routing in our overlay network and consequently increase the amount of monitorability or anonymity, we enforce a randomized message sending scheme similar to the one proposed by Valiant \cite{valiant_scheme_1982}. The combination of enforced message sending scheme and spectral properties of the overlay network graph allow each monitor to supervise $\epsilon$-close fraction of the network traffic. In case of anonymity network, random routing increases unpredictability in the network, which is inevitable for high anonymity.

The monitoring of the network resorts to monitors that are located at the nodes of the constructed overlay network. Monitors are not required to be located at each node of the network but can be placed at selected nodes. This decision depends on the selected message sending policy of the network. The combination of enforced message sending scheme and spectral properties of the overlay network graph provides monitoring abilities such that each monitor supervises $\epsilon$-close fraction of the network traffic. Uniquely, under certain conditions, our overlay network construction provides a uniform dispersion of flows in the overlay network. In such case, all monitors will be equally exposed to the network traffic. In case of anonymity network, this reduces over-exposure of anonymity relay node, and thus, reduces centrality of relay nodes.




By its very nature, the constructed network is robust and is able to recover from congestion and link failures due to expansion properties, and due to path diversity which is achievable by a combination of random spanning trees \cite{goyal_expanders_2009,motiwala_path_2008}.

Extended abstract of this work appeared as a conference paper \cite{dolev_monitorability_2017} which presented the preliminary results. In this work, we provide full proofs and detailed discussions. 

Next, we discuss the related works. Later, we describe the overlay network construction method and distributed expansion verification algorithm used in a distributed construction of the overlay network. In Sections \ref{sec:exp_graph_mons} and \ref{sec:measure_mons_sccs} we describe the use of expander graphs for monitoring, the message sending scheme, and the probabilities for successful network monitoring. Finally, in Section \ref{sec:anonymity} we show a feasible implementation of our on-demand network construction as an on-demand anonymity service, which we believe opens a new scope in the research on communication anonymity. 

\subsection{Related Work}

Goyal et al. demonstrated the building process of an expander graph, whose expansion depends on the degree of the primal graph, by the union of random spanning trees and showed that the union of the trees approximates each cut of the primal graph within a factor of $O(\log n)$. In our work, we show that the same method can be used for building (weighted) expanders from a weighted graph. 

Spielman and Teng introduced the notion of spectral sparsification \cite{spielman_spectral_2011}, deriving a stronger notion than cut sparsification introduced by Bencz\'ur and Karger in \cite{benczur_approximating_1996}, and constructed sparsifiers with $O(n \log ^c n)$ edges, where $c$ is a large constant. Later in \cite{spielman_graph_2011}, Spielman and Srivastava presented a nearly-linear time algorithm for graph sparsifiers with $O(n \log n / \epsilon^2)$ edges. After, their result was improved in \cite{batson_twice-ramanujan_2012} proving that every graph has a spectral sparsifier with a number of edges linear in its number of vertices. We show that our method of construction is equivalent to the method used in \cite{spielman_graph_2011} for graph sparsification. 

Similar result to ours was shown by Fung et al. in \cite{fung_general_2011}. Fung et al. showed that sampling spanning trees while adjusting their link weights results in a cut sparsifier. In contrast to their work, we utilize the weights of the given primal graph using a random walk in order to construct a capacity optimized expander, while including enough edges results in a spectral sparsifier of the primal graph. Besides, performing a random walk on a weighted graph better preserves the locality\footnote{We thank Noga Alon for drawing our attention to this observation.} of a cut. 

Additionally, our construction method can run in a distributed manner concurrently constructing the expander graph. Distributed construction of expander graphs was also shown by Dolev and Tzachar in \cite{dolev_spanders:_2010}, where the authors introduced the notion of {\it Spanders}, distributed spanning expanders, and showed a practical way for verifying that the constructed graph is an expander. We use the method of expansion verification presented in \cite{dolev_spanders:_2010} in order to optimize the construction of the overlay network and limit the number of constructed spanning trees.

We show that our network architecture is valuable for network monitoring. Measures for estimating monitoring capabilities of a vertex (BC, SPBC, LC, FBC) were established in \cite{freeman_set_1977,goh_universal_2001,borgatti_centrality_2005,freeman_centrality_1991}. Routing betweenness centrality (RBC) is a network measure, which was proposed in \cite{dolev_routing_2010}, for estimating the control probabilities of a vertex or a set of vertices. RBC generalizes aforementioned network measures and estimates the extent to which vertices or group of vertices are exposed to network traffic~\cite{dolev_routing_2010}. Consequently, RBC is useful for predicting the effectiveness (and cost) of passive network monitoring. Altshuler et al. showed an efficient flooding scheme for generating a collaboration between a group of random walking agents which are released from different sources and at different times \cite{altshuler_ttled_2012}. This participation of agents results in a collaborative monitoring infrastructure, requiring only a small number of active monitors. We show that our network construction requires only a few monitors, dependent on the message policy, for monitoring the whole network with high probability. While if utilizing a particular enhancement of the construction, using the RBC measure, we show that each of the monitors in our constructed network is equivalently effective as a passive monitor.

Cryptographic primitives are used to establish a secure communication channel and ensure confidentiality, provide digital signatures, message authentication codes, and hash functions. However, cryptographic primitives cannot provide anonymous communication since the channel is susceptible to traffic analysis. Encrypted messages can be tracked and reveal the two communicating parties. The information with regard to the identity of the communicating parties can be sensitive. 

An extensive amount of work was done on the subject of anonymity networks and anonymity protocols. Chaum introduced the concept of a {\itshape mix} \cite{chaum_untraceable_1981} and mix-net based protocols, and also a DC-net \cite{chaum_dining_1988}, a broadcast network which provides both sender and receiver anonymity. DC-net scheme suffers from poor scalability and it is unsuitable for large-scale networks \cite{hermoni_rendezvous_2014}. Most notable work based on DC-net is Xor-Trees \cite{dolev_xor-trees_2000}, which was proposed by Dolev and Ostrovsky to provide sender and receiver anonymity, and for also, reducing the amount of communication overhead. Additional anonymous communication scheme which provides a high degree of anonymity is Buses \cite{beimel_buses_2002}, which is a network routing based anonymous communication scheme, that can be viewed as a bus system. Buses attempts to hide traffic patterns and to provide an unlinkability for two communicating parties.
Recent work by Hermoni et al. proposed a Peer-to-Peer file sharing system which provides anonymity to all participants, namely, receiver (server) and sender (publisher or reader) anonymity~\cite{hermoni_rendezvous_2014}. Hermoni et al. propose the use of anonymity tunnels for each different user. 

Network topology can play an important factor in the anonymity provided by the anonymity service. The authors of \cite{diaz_impact_2010} and \cite{danezis_mix-networks_2003} show that restricted network topologies provide better anonymity, less cover traffic overhead, and scale better as the number of (mix) nodes grows. Additionally, Danezis proposes the use of expander graphs based topology for anonymity networks and shows that utilizing properties of expander graphs can provide scalable and highly anonymized networks. We provide a practical construction of expander-based topology, and further, show advantages of sparse overlay network with expansion properties and its consequent gain for anonymity on-demand services. 

Our results can be used as a basis for providing a flexible and robust network architecture as a service with on-demand deployment. Boubendir describes an implementation of NaaS architecture with SDN-enabled NFV in \cite{boubendir_naas_2016}, and shows feasible on-demand dynamic network service based on SDN-enabled NFV \cite{boubendir_-demand_2016}. We further exploit the NV, SDN and NFV emerging technologies to enable network architecture as a service for use in private commercial networks, network and service providers, or facilities desiring flexible policy enabled networking to secure their traffic and to monitor network flows for mitigation of misuse or malicious uses.

\subsection{Results}
Our results can be summarized as follows. We generalize the work in \cite{goyal_expanders_2009, frieze_expanders_2014} and show that it is possible to construct an expander graph via random spanning trees of a weighted graph, obtaining a sparse construction which can be used for building capacity oriented overlay networks with expansion close to the substrate graph within a factor of $\log n$. Generating enough random spanning trees will result in a sparsifier, and as described later, generation of random spanning trees via weighted random walk results in a spectral sparsifier. 

We use the union of random spanning trees for constructing an overlay network with expansion properties. Augmenting the overlay network with a randomized routing method similar to Valiant's, we provide a construction with optimal monitoring abilities. With additional enhancement we can achieve uniform dispersion of network flows, deriving a maximal entropy in the system. Subsequently, our system can be used as an anonymity network. Finally, utilizing SDN-based infrastructure our overlay network can be (re)constructed on-demand, and thus, mitigate certain attacks on anonymity networks.

\subsection{Preliminaries}
Throughout the work we assume that $G=(V, E, \omega)$ is a weighted, connected, and undirected graph having a vertex set $V=\{1,\dots,n\}$, an edge set $E\subseteq\{(u,v)|\ u, v\in V\}$, and a weight function $\omega:E\rightarrow\mathbb{R}^+$. For each vertex $u \in V$, $\omega_u=\sum_{z}\omega_{(u,z)}$ is the total weight of edges that are incident to the vertex $u$. The maximum degree of a graph is denoted by $\Delta_G$, and the minimum degree is denoted by $\delta_G$ (the subscript will be omitted if the particular graph is understood from the context).
For $S, T \subset V$, we specify the set of edges emerging from $S$ to $T$ by 
$$E(S, T)=\{(u,v) \mid u \in S, v \in T, (u,v) \in E\}$$
We denote the \textit{edge boundary} of a set $S$ as $\partial S$ and is defined as $\partial S=E(S, \bar{S})$. The edge boundary is a set of edges emerging from the set $S$ to its complement. We specify the set of neighbors of $v \in V$ as $\Gamma(v)=\{u \in V | (u,v) \in E\}$. For $A \subseteq V$, $\Gamma(A)=\cup_{v \in A} \Gamma(v)$ and $\Gamma'(A)=\Gamma(A) \setminus A$. 

\paragraph{Graph Expansion.}
We provide only the basic definitions of expander graphs. For a good introduction to expander graphs see \cite{hoory_expander_2006,goldreich_basic_2011}. Expansion requires that any set of vertices, of size at most $n/2$, has a relatively large set of neighbors.
The edge expansion ratio of $G$ is defined as:

\[
\gamma^E(G)=\min_{\{S \subseteq V: \abs{S} \leq \frac{n}{2}\}} \frac{\abs{\partial S}}{\abs{S}}
\]
The vertex expansion of graph $G$ is defined as:

\[
\gamma^V(G)=\min_{\{S \subseteq V: \abs{S} \leq \frac{n}{2}\}} \frac{\abs{\Gamma'(S)}}{\abs{S}}
\]
\paragraph{The Laplacian.}
The \textit{Laplacian } matrix of a weighted graph $G=(V, E, W)$ is defined by
\[
L(u,v)= \begin{cases}
-\omega_{(u,v)}, & \text{if $u \neq v$} \\
\sum_z \omega_{(u,z)}, & \text{if $u = v$} 
\end{cases}
= D - A
\]
where $A$ is the weighted adjacency matrix
\[
A(u,v) = \begin{cases}
\omega_{(u,v)}, & u \neq v \\
0, & u = v
\end{cases}
\]
and $D$ is the diagonal matrix of the weighted degrees
\[
D(u,v) = \begin{cases}
\omega_{u}, & u = v \\
0, & u \neq v
\end{cases}
\]

We can orient the edges of $G$ arbitrarily. For an edge $e=(u,v)$, arbitrarily oriented from $u$ to $v$, define $$b^T_{e=(u,v)}=\textbf{1}_v-\textbf{1}_u$$ where $\textbf{1}_u$ is the elementary unit vector defined as
\[
\textbf{1}_u(v)=\begin{cases}
1 & \text{if $u=v$}, \\
0 & \text{otherwise}
\end{cases}
\]
We define the \textit{edge-vertex incidence} matrix, for $|V|=n$ and $|E|=m$,
\[
B_{m\times n}=
\begin{pmatrix}
b^T_{e_1}\\ 
b^T_{e_2}\\ 
\vdots\\ 
b^T_{e_m}\\ 
\end{pmatrix}
\]
It is possible to define then the Laplacian of an edge $e=(u,v)$ as $$L_{e}=\omega_eb^T_eb_e$$
and the Laplacian of a graph $G$ can be defined as $$L=\sum_{e \in E}\omega_eb^T_eb_e=B^T W B$$
where $W_{m \times m}$ is a diagonal matrix of weights, $W(e,e)=\omega_e$.
The quadratic form associated with $L$ takes the value
\begin{align}
\label{eq:qform}
\forall x \in \mathbb{R}^n: \ x^TLx&=x^T B^T W Bx \\
\nonumber &=\Vert W^{1/2}Bx\Vert^2_2 \\
\nonumber &=\sum_{(u,v) \in E} \omega_{(u,v)}(x(u)-x(v))^2 \geq 0
\end{align}
%
The quadratic form measures the smoothness of the function that gives values of the vector $x \in\mathbb{R}^n$ to the vertices of the graph. 
The matrix $L$ is symmetric and we can diagonalize it by writing $$L=\sum_{i=1}^{n-1}\lambda_i u^T_i u_i$$ where $\lambda_1, \ldots, \lambda_{n-1}$ are the nonzero eigenvalues of $L$ and $u_1, \ldots, u_{n-1}$ are the corresponding set of eigenvectors. 

The Laplacian matrix is not invertible since it has $0$ as an eigenvalue. We will be interested in the Moore-Penrose Pseudoinverse of $L$, denoted as $L^+$, given by
\[
L^+=\sum_{i=1}^{n-1}\frac{1}{\lambda_i}u^T_i u_i
\]

\paragraph{Effective Resistance.}
We can represent an electrical network with a graph $G$, treating each edge $e$ as a resistor with \textit{conductance} $\omega_e$. The effective resistance between a pair of nodes $u$ and $v$, denoted as $R_{(u,v)}$, is the electrical resistance measured across the nodes $u$ and $v$. In other words, if we define a potential function $\phi:V \rightarrow \mathbb{R}$ along the vertices of the graph, and $\textbf{i}(e)$ is an electrical flow along the edge $e$. According to Ohm's law, for any edge $e=(u,v)$, 
\begin{equation*}
\label{eq:ohm}		
\textbf{i}(u,v)=\omega_e(\phi(u) - \phi(v))
\end{equation*}
Assume that we inject one unit of current at $u$ and measure it at $v$. By Kirchoff's current conservation law 
\begin{equation*}
\label{eq:conserve}
B^T\textbf{i}=b^T_{(u,v)}
\end{equation*}
Combining the above, we obtain 
\begin{equation*}
b^T_{(u,v)}=B^T W B \phi=L\phi
\end{equation*}
or using the pseudoinverse Laplacian matrix
\begin{equation*}
L^+b^T_{(u,v)}=L^+L\phi=\phi
\end{equation*}
Multiplying by $b_{(u,v)}$ on the left we obtain the potential difference across the edge $e=(u,v)$: $$b_{(u,v)}L^+b^T_{(u,v)}=b_{(u,v)}\phi=\phi(v) - \phi(u)$$
The quantity $R_{e=(u,v)} = b_{(u,v)}L^+b^T_{(u,v)}$ is the effective resistance across the edge $e$. Further, we define the matrix $\Pi=W^{1/2}BL^+B^TW^{1/2}$ that has as its diagonal the entries 
\begin{equation*}
\Pi(e,e)=\sqrt{W(e,e)}R_e\sqrt{W(e,e)}=\omega_eR_e
\end{equation*}
For any pair of edges $e$ and $f$
\begin{equation*}
\Pi(e,f)=\sqrt{W(e,f)}R_e\sqrt{W(e,f)}=\sqrt{\omega_e \omega_f}b_{e}L^+b^T_{f}=\sqrt{\omega_e \omega_f}R_{ef}
\end{equation*}

\paragraph{Spectral Graph Sparsification.}
Let $G=(V, W, \omega)$ be a weighted undirected graph and $H=(V, \tilde{E}, \tilde{\omega})$ a weighted undirected graph on the same set of vertices. Let $L_G$ be the Laplacian matrix of $G$ and $L_H$ the Laplacian matrix of $H$. Given $\epsilon > 0$, $H$ spectrally sparsifies $G$ if $\forall x \in \mathbb{R}^n$

\begin{equation}
\label{eq:laplace_preserve}
(1-\epsilon)x^TL_Gx \leq x^TL_Hx \leq (1+\epsilon)x^TL_Gx
\end{equation}

\section{Construction of Expander-Based Overlay Network}
\label{sec:alg_desc}
The substrate network is represented by a weighted graph $G=(V,E,\omega)$. We construct an overlay network with expansion properties as a union of random spanning trees, each tree constructed with a \textbf{weighted random walk} on the substrate graph. A weighted random walk begins from a randomly chosen vertex $u$, and moves to one of its neighbors $v \in \Gamma'(u)$ with probability proportional to the weight of a connecting edge $(u,v)$: $Pr[(u,v)]=\frac{\omega(u,v)}{\omega(u)}$. Each time the random walker arrives at a new vertex, a vertex which was not visited before, the edge through which it arrived is added to the spanning tree construction. 

Algorithm \ref{alg:alg1} generates a random spanning tree using a modified version of the algorithm derived by Andrei Broder~\cite{broder_generating_1989}. Algorithm \ref{alg:alg2} creates an overlay network by repeating $k$ times the creation of a random spanning tree using Algorithm \ref{alg:alg1}, and uniting the generated random spanning trees. The weights of the added edges are re-scaled by the total number of samples and the re-scaled weights are summed if edges already exist.

\begin{algorithm}
	\small
	\caption{\small Generation of Random Spanning Tree via Random Walk Simulation}
	\label{alg:alg1}
	\Input{$G=(V,E, \omega)$} 
	\Output{spanning tree $T$} 
	\BlankLine
	$T$ = \{\} \\
	Simulate a weighted random walk on graph $G$ starting at an arbitrary vertex $s$ until every vertex is visited. 
	$T \leftarrow T \cup e(v,u)$ for each vertex $u\in V$ \textbackslash   $~s$ if this is the first visit of vertex $u$. \\
	\Return $T$
\end{algorithm}

The running time of Algorithm \ref{alg:alg1} depends on the cover time of the graph. It is known that the cover time of a graph is captured by the effective resistance of the graph and is bounded by $\bigO(2|E|R_{\max}\log n)$, where $R_{\max}$ is the maximal effective resistance of the graph~\cite{chandra_electrical_1996}. By treating the graph as an electrical network we replace $|E|$ by the sum of all resistances $\mathcal{R}=\sum_{e \in E}R_e=\sum_{e \in E}\frac{1}{\omega_e}$. There, the cover time bound is $\bigO(2\mathcal{R}R_{\max}\log n)$. The maximal effective resistance is bounded by $\bigO(n)$~\cite{chandra_electrical_1996}. Consequently, the cover time is bounded by $\bigO(2\mathcal{R}n\log n)$. Note that for very small weights, the cover time can be exponentially large. Since the random walk will be very unlikely to go across a cut with all edges having exceptionally small weights. In general, the cover time of a weighted random walk is $\Omega(n \log n)$~\cite{berenbrink_random_2015}.

The running time of Algorithm \ref{alg:alg2} is $k$ times the running time of Algorithm \ref{alg:alg1}, which would be at most $\bigO(2k\mathcal{R}n\log n)$ and at least $\Omega(kn \log n)$.

\begin{algorithm}
	\small
	\caption{\small Expander Overlay Construction}
	\label{alg:alg2}
	\DontPrintSemicolon
	
	\Input{$G=(V,E, \omega)$, $k$}
	\Output{union of $k$ random spanning trees $U_{G}^{k}$}
	\BlankLine
	$U_{G}^{k}=\{\}$ \\
	\RepTimes{k} {
		Generate a random spanning tree $T$ according to Algorithm \ref{alg:alg1}. \\
		$U_{G}^{k} \leftarrow U_{G}^{k} \cup T$ \\
	}
	\Return $U_{G}^{k}$
\end{algorithm}
Since our construction is based on random spanning trees, we need to take into account the probabilities related to the edges of a random spanning tree. First, we show that in a weighted graph the probabilities for two edges to belong to a random spanning tree are negatively correlated.

\begin{lemma}
	\label{thm_neg_corr}
	For any weighted, undirected, connected graph $G$, let $T$ be a random spanning tree. If $e$ and $f$ are distinct edges, then $$Pr[e,f \in T] \leq Pr[e \in T]Pr[f \in T]$$
\end{lemma}

\begin{proof}
	It is well established (see \cite{snell_topics_1995,batson_spectral_2013} for details) that if $T$ is a random spanning tree, then the probability that an edge $e$ belongs to $T$ is $P[e \in T]=\omega_e b_{e}L^+b^T_{e}=\omega_eR_e=\Pi(e,e)$ and for a pair of edges $e$ and $f$ 
	
	\begin{align*}
	Pr[e,f \in T]&=\det \begin{pmatrix}
	\Pi(e,e) & \Pi(e,f) \\
	\Pi(e, f) & \Pi(f, f) 
	\end{pmatrix} \\
	&=\Pi(e,e)\Pi(f,f)-\Pi(e,f)\Pi(e,f) \\
	&=\Pi(e,e)\Pi(f,f)-\left(\Pi(e,f)\right)^2 \\
	\end{align*}
	Since $\left(\Pi(e,f)\right)^2 \geq 0$, we obtain that $Pr[e,f \in T] \leq Pr[e \in T]Pr[f \in T]$
	
\end{proof}
\textbf{Negative Correlation of Edges.} Lemma \ref{thm_neg_corr} can be extended \cite{snell_topics_1995, goyal_expanders_2009, frieze_expanders_2014}
for any subset of edges $e_1, \dots,e_k \in E$:
\begin{equation}
\label{eq:neg_corr_edges}
Pr[(e_1 \in T), \dots, (e_k \in T)] \leq Pr[e_1 \in T] \cdots Pr[e_k \in T]
\end{equation}
or similarly for the complement event
\begin{equation*}
\label{eq:neg_corr_edges_complement}
Pr[(e_1 \notin T), \dots, (e_k \notin T)] \leq Pr[e_1 \notin T] \cdots Pr[e_k \notin T]
\end{equation*}
Accordingly, we define the following indicator variables:
\[
X_e = \left\{\begin{matrix}
1 & e \in T \\ 
0 & otherwise
\end{matrix}\right.
\]
Now, we can rewrite \eqref{eq:neg_corr_edges} as
\begin{equation*}
\label{eq:neg_corr_rand_var}
E[X_{e_1} \cdots X_{e_k}] \leq E[X_{e_1}] \cdots E[X_{e_k}]
\end{equation*}
%
It is possible to apply Chernoff bounds unaltered to negatively correlated variables~\cite[Proposition 5]{dubhashi_balls_1998}, and by \cite[Theorem A.1.13]{alon_probabilistic_2004} it is possible to derive the following version of Chernoff bounds \cite{goyal_expanders_2009}:

\begin{theorem}
	\label{thm_neg_corr_chernof}
	Let $\{X_i\}_{i=1}^n$ be a family of \textup{$0$--$1$} negatively correlated random variables such that $\{1-X_i\}_{i=1}^n$ are also negatively correlated. Let $p_i$ be the probability that $X_i=1$. Let $p=\frac{1}{n}\sum_{i \in [n]}p_i$. Then for $\lambda > 0$ 
	$$Pr[\sum_{i \in [n]}X_i < pn - \lambda] \leq e^{\frac{-\lambda^2}{2pn}}$$
\end{theorem}
\paragraph{\textbf{Base graph is a complete graph.}} In this part, we show that the union of random spanning trees is a vertex expander of the complete graph when at least two random spanning tree are required. In a complete graph, we can treat the edge selection through a random walk as a random neighbor selection. However, we use negative correlation property in order to take into account the not completely independent selection. We show that two random spanning trees are required in order to approximate the expansion of a complete weighted graph.
\begin{theorem}
	\label{thm_expansion_complete_g}
	The union of two random spanning trees of the complete undirected weighted graph $G$ on $n$ vertices has a constant vertex expansion with probability $1-o(1)$.	
\end{theorem}
\begin{proof}
	Each random spanning tree is generated by a weighted random walk. The process repeats for $k$ times obtaining $k$ random spanning trees. Those trees are then combined into a graph $U^k_G$.
	
	Using the probabilistic method we show that in a complete undirected weighted graph the probability of the union of random spanning trees not having a constant expansion is small. We will prove this by showing that the complement event (i.e., that there is no such expansion of $G$) has a probability approaching $0$. For a set $S \subseteq V$ of size $s \leq n/2$ and an expansion constant $\gamma$, we will upper bound the probability that $$Pr[\frac{\abs{\Gamma'_{U^k_G}(S)}}{\abs{S}} \leq \gamma ] $$ and then union bound for all possible sets $S$.
	%
	%
	%
	%
	%
	As we saw earlier, the probability of an edge $e$ to belong to a random spanning tree $T$ is given by $p_e=Pr[e \in T]=\omega_eR_e$, while the complementary probability is $Pr[e \notin T]=(1-p_e)$.
	%
	Negative correlation implies that the probability for all edges of the edge boundary to not belong to a random spanning tree is given by 
	$$Pr[\forall e \in E(S, \bar{S}):e \notin T]=\mathop{\prod}_{e \in E(S, \bar{S})}(1-p_e)$$
	%
	Therefore, the probability of the expansion of a union of random spanning trees is less than $\gamma$ is given by
	\begin{equation*}
	\begin{aligned}
	Pr[\min_{\{S \subseteq V: |S| \leq \frac{n}{2}\}}\frac{\abs{\Gamma'_{U^k_G}(S)}}{\abs{S}} < \gamma] & \leq 
	\sum_{s=1}^{\frac{n}{2}} \binom{n}{s} \binom{n}{\gamma s} 
	\left(
	\mathop{\prod}_{e \in E(S, \bar{S})} (1-p_e) 
	\right) ^ {ks} \\
	& \leq \sum_{s=1}^{\frac{n}{2}} 
	\left( \frac{ne}{s} \right)^s \left( \frac{ne}{\gamma s}  \right)^{\gamma s} 
	\left(
	\mathop{\prod}_{e \in E(S, \bar{S})} (1-p_e) 
	\right) ^ {ks} \\
	& \leq 
	\sum_{s=1}^{\frac{n}{2}} 
	\left[
	\left( \frac{n}{s} \right) 
	\left( \frac{n}{\gamma} \right) ^ \gamma
	\cdot \left(\frac{e}{\gamma}\right)^\gamma \cdot e 
	\cdot 
	\left(
	\mathop{\prod}_{e \in E(S, \bar{S})} (1-p_e) 
	\right)^k 
	\right] ^ {s} \\
	\end{aligned}
	\end{equation*}
	%
	%
	Since a spanning tree has $n-1$ edges then
	\begin{align*}
	\mathop{\prod}_{e \in E(S, \bar{S})} (1-p_e) & \leq (1-p_e)^n
	\end{align*}
	%
	We can always find $z \in \mathbb{Z}^+$, so that $(1-p_e)^n \leq \frac{z}{n}$. It follows then
	\begin{align*}	
	Pr[\min_{\{S \subseteq V: |S| \leq \frac{n}{2}\}}\frac{\abs{\Gamma'_{U^k_G}(S)}}{\abs{S}} < \gamma] &\leq
	\sum_{a=1}^{\frac{n}{2}} 
	\left[
	\left( \frac{n}{s} \right) 
	\left( \frac{n}{s} \right) ^ \gamma
	\cdot \left(\frac{e}{\gamma}\right)^\gamma \cdot e 
	\cdot 
	\left(\frac{z}{n}\right)^{k}
	\right] ^ {s} \\ 
	&\leq \cdot \sum_{s=1}^{\frac{n}{2}} \cdot 
	\left[
	\left(\frac{1}{n}\right)^{k-\gamma-1} \cdot e^{\gamma+1} \cdot z^k \cdot \left(\frac{1}{s}\right)^\gamma
	\right]^s =
	o(1) 
	\end{align*}
	%
	For $k \geq 2$ and sufficiently small $\gamma$ this sum goes to zero as $n\rightarrow\infty$. 
\end{proof}

It is important to note here, that the expansion due to two random spanning trees can be very low. Consider for example, a complete graph and a balanced cut, with $n/2$ edges on each side, so that one edge across the cut has a weight that is much greater than the weight of the other edges across the cut. In such a graph, it is extremely likely that in majority of times, if not always, only the edge with the greatest weight will belong to a random spanning tree. The resulting union of two random spanning trees in such case, will still be an expander graph, just a very \textit{bad} one. More trees will be required in order to improve the expansion.


\paragraph{\textbf{Base graph is a bounded-degree graph.}} Now, we consider a weighted, undirected, irregular bounded-degree graphs, in which the maximum degree $\Delta$ is a constant independent of $n$. We show that for such graphs the union of random spanning trees captures the edge expansion of the underlying graph. 

\begin{theorem}
	\label{thm:expansion_bounded_g}
	For an undirected, irregular, and weighted graph $G=(V,E, \omega)$, let $U_G^k$ denote the union of $k$ random spanning trees of $G$. Also, let $\alpha > 0$ be a constant and 
	$\alpha(k-1) > \frac{8}{p}$, where $p$ is an average probability for edge $e$ being part of a random spanning tree. Then with probability $1-o(1)$, for every $A \subseteq V$,  $\abs{A} \leq \frac{n}{2}$, we have
	$$\abs{\partial_{U_G^k}A} \geq \frac{1}{\alpha \ln (n)} \abs{\partial_GA}$$
\end{theorem}

\begin{proof}
	We follow the proof in \cite{goyal_expanders_2009} while modifying the parts related to the weighted characteristics of the graph. 
	As we have already shown, the probability for an edge $e$ being part of a random spanning tree is
	\[
	Pr[e \in T] = \Pi(e, e)
	\]
	whereas the average probability is 
	\[
	p=\frac{\sum_{e \in E} Pr[e \in T] }{\abs{E}}
	\]
	Therefore, for $A \subseteq V$ $$E[\abs{\partial_{T}A}] = p \abs{\partial_GA}$$
	We have $\abs{\partial_TA}=\sum_{e \in \partial_GA}X_e$. Since the random variables $\{X_e\}$ are negatively correlated, we use Theorem 
	\ref{thm_neg_corr_chernof}:
	\[
	Pr[\sum_{e \in \partial_GA} X_e < p\abs{\partial_GA} - \lambda] < e^{-\lambda^2/(2p\abs{\partial_GA})}
	\]
	where the average of $Pr[X_e=1]$ is $p$. For $\lambda=\frac{1}{2}p\abs{\partial_GA}$ we have
	\begin{equation*}
	Pr [\abs{\partial_{T}A} < \frac{1}{2}p\abs{\partial_GA} ] < e^{-\frac{1}{8}p\abs{\partial_GA}}
	\end{equation*}
	which gives the following bound for $k$ trees
	\begin{equation}
	\label{eq:ugk_bound}
	Pr[|\partial_{U_G^k}A| < \frac{1}{2}p\abs{\partial_GA} ] < e^{\frac{-k}{8}p\abs{\partial_GA}}
	\end{equation}
	The expected number of sampled edges in a cut must be at least $\ln(n)$ \cite{karger_random_1994}. There, to have an expansion property in the union of random spanning trees, those cuts must be greater than their size in $G$. We would like to estimate the probability that there is a \textbf{bad} cut. A cut whose edge expansion in the union of random spanning trees is less than in the underlying graph $G$.
	More formally, a cut $A$ is bad if $\abs{\partial_{U_G^k}A}=a$ and $\abs{\partial_GA} \geq \alpha a \ln(n)$. Such a cut must have lesser edge expansion in all remaining trees. We use (\ref{eq:ugk_bound}) for estimating the occurrence of such event. The number of cuts of size $a$ in the first tree is at most $\binom{n-1}{a}$. The probability that a bad cut exists
	
	\begin{equation*}
	\begin{aligned}
	& \sum_{a=1}^{n / \ln (n)} \binom{n-1}{a} e^{-(k-1)\alpha a \ln (n) \frac{1}{8} p } \\ 
	& < \sum_{a=1}^{n / \ln (n)} \binom{n}{a} e^{-(k-1)\alpha a \ln (n) \frac{1}{8} p } \\ 
	& \leq \sum_{a=1}^{n / \ln (n)} \left(\frac{ne}{a}\right)^a e^{-(k-1) \alpha a \ln (n) \frac{p}{8}} \\ 
	& = \sum_{a=1}^{n / \ln (n)} exp \left( \ln \left(\frac{ne}{a}\right)^a + \ln \left(e^{-(k-1) \alpha a \ln (n) \frac{p}{8}}\right) \right) \\
	& = \sum_{a=1}^{n / \ln (n)} exp \left ( a \left( \ln \left(\frac{e}{a} \right) + \ln (n) \left( 1 - (k-1)\alpha \frac{p}{8}\right) \right) \right)
	\end{aligned}
	\end{equation*}
	Choosing $(k-1)\alpha > \frac{8}{p}$ makes the above sum $o(1)$.
\end{proof}

The probability of the random walker moving to the closest neighbor depends on the degree of the graph. For  unweighted graphs, let $\bar{d}$ be the average degree of the graph, while in the general case $\bar{d}$ is the average weighted degree. Then, the average probability $\bar{p} = \frac{1}{\bar{d}}$ and so $ k $ on average is at least $ \frac{8\bar{d}}{\alpha} $. 

\paragraph{\textbf{$\pmb{U_G^k}$ is a sparsifier of a bounded-degree graph.}} We show that $U_G^k$ generated by Algorithm~\ref{alg:sparsify} is a sparsifier of the weighted graph $G$, if there is a sufficient number of generated random trees.

\begin{algorithm}
	\small
	\caption{\small Sparsify}
	\label{alg:sparsify}
	\DontPrintSemicolon
	
	\Input{$G=(V, E, \omega$)}
	\Output{$U^k_G=(V, \tilde{E}, \tilde{\omega})$}
	
	\BlankLine
	$U^k_G = \{\}$ \\
	$i = 1$ \\
	
	\Repeat{$k$ trees are generated}
	{
		Generate a random spanning tree $T$, according to Algorithm \ref{alg:alg1} \\
		\ForEach{$e \in E(T)$}
		{
			$p_e = \frac{Pr[e \in T]}{\sum_{e'} Pr[e' \in T]}$\\		
			Add edge $e$ with the weight $ \omega_e / ip_e$ to $U^k_G$, summing weights if $e$ already exists in $U^k_G$ \\
			$i \leftarrow i + 1$ \\
		}
		
	}
\end{algorithm}

\begin{theorem}
	Let $L$ be the Laplacian matrix of weighted, undirected, bounded degree graph $G$ and $\tilde{L}$ the Laplacian matrix of $U_G^k$. Let $C n \log n \leq k \leq C n \log n / \epsilon^2$,  where $\frac{1}{\sqrt{n}} < \epsilon \leq 1$ and $C > 0$. Then with a probability of at least $\frac{1}{2}$
	
	$$ \forall x \in \mathbb{R}^n \ \ \ (1-\epsilon)x^{T}Lx \leq x^{T}\tilde{L}x \leq (1+\epsilon)x^{T}Lx $$
\end{theorem}
\begin{proof}
	
	We prove this by showing that the process of creating $U^k_G$ is similar to sampling with probability proportional to the effective resistance. We use the same notation used in \cite{spielman_graph_2011} for similar mathematical structures. 
	
	The matrix $S_{m \times m}$ of $U^k_G$ is a nonnegative diagonal matrix with a random entry $S(e,e)$ that specifies the normalized number of times edge $e$ was included in $U^k_G$:
	\[
	S(e,e)=\frac{\text{\# of times $e\in T$}}{\abs{U^k_G}p_e} = \frac{\tilde{\omega_e}}{\omega_e}
	\]
	The weight of $e$ in $U^k_G$ is given by $\tilde{\omega_e}=S(e,e)\omega_e$. The Laplacian $\tilde{L}$ of $U^k_G$ can be written as 
	\[
	\tilde{L} = B^T \tilde{W}B = B^TW^{1/2}SW^{1/2}B
	\]
	The probability that an edge $e$ belongs to a random spanning tree is equal to $\Pi(e,e)$, and the amount of edge $e$ in the approximated graph is $\tilde{\Pi}(e,e)$. We need to show that $\lVert \tilde{\Pi} - \Pi \rVert_2$ is small. Since $\Pi$ is a projection matrix \cite{spielman_spectral_2011}, then $\Pi^2=\Pi$.
	According to Lemma~$4$ from \cite{spielman_graph_2011}, if $\lVert \Pi S \Pi - \Pi \Pi \rVert_2 \leq \epsilon$ then 
	\[
	\forall x \in \mathbb{R}^n~~(1-\epsilon)x^TLx \leq x^T\tilde{L}x\leq (1+\epsilon)x^TLx
	\]
	This says that if we preserve $\Pi$ well enough then the Laplacian is preserved too. The concentration result in \cite[Theorem 3.1]{rudelson_sampling_2007} and \cite[Lemma 5]{spielman_graph_2011} show that if we have $\Omega(n \log n )$ samples, then with a probability of at least $1/2$ 
	\[
	\lVert \Pi S \Pi - \Pi \Pi \rVert_2 \leq \epsilon
	\] 
	Since each spanning tree is of size $n-1$, then $\Omega( \log n )$ trees are required in order to sparsify~$G$.
	%
	%
	%
	Whereas, for the concentration result from \cite{rudelson_sampling_2007} to hold for a small error term $\epsilon$, Spielman and Srivastava show in \cite{spielman_graph_2011} that we need $\bigO(n \log n / \epsilon^2)$ samples, where $\frac{1}{\sqrt{n}} < \epsilon \leq 1$. Thus, we need $\bigO(\log n /\epsilon^2)$ trees for the union of random spanning trees to sparsify. 
	%
	%
	
\end{proof}

Spectral sparsification tells us that for any real function on the vertices, the weight of each cut of the overlay network is approximately close to the weight of each cut of the substrate network. Assuming that the weight $\omega_e$ represents the amount of traffic passing through a link $e$, we obtain a remarkable result. The amount of traffic sampled through a cut of the overlay network is $(1 \pm \epsilon)$ close to the amount of traffic sampled through a cut of the substrate network. In other words, a monitor located at one of the nodes of the overlay network can sample and monitor an $\epsilon$-close amount of traffic.

It is important to note that, weighted generation of random spanning trees plays a vital role in the algorithm. If we were sampling the trees uniformly, we would not be able to obtain a spectral sparsifier. As an example, consider a weighted graph with one edge having a much greater weight than that of the other edges. This edge plays an essential role in the resulting sparsifier. Thus, including this edge sufficiently in the union of random spanning trees in order to correctly approximate its weight is essential for preserving the quadratic form of the graph and the weight of a cut in which this edge participates (as can be seen from Equations~(\ref{eq:qform}) and~(\ref{eq:laplace_preserve})). Uniformly choosing random spanning trees will not result in a spectral sparsifier but only in a cut sparsifier \cite{batson_spectral_2013, fung_general_2011, fung_graph_2010} since there is an actual probability that the edge with the greater weight will not be included in a uniformly chosen random spanning tree.

The running time of Algorithm \ref{alg:sparsify} follows from running time analysis of Algorithm~\ref{alg:alg2} in which we substitute our result for $k$ and obtain the running time $\bigO(2\mathcal{R}n\log^2 n / \epsilon^2)$, and the bounded below running time $\Omega(n \log^2 n / \epsilon^2)$.
	


\subsection{Distributed Construction of the Overlay Network}
The algorithm for overlay network construction (Algorithm \ref{alg:alg2}) can be modified to a distributed algorithm. This is equally beneficial for networks where a centralized entity consists of several cores that can execute the algorithm in a distributed manner. For example, in software-defined networking, the network control software can execute the iterations of the algorithm concurrently or delegate the execution of some iteration to a different controller. Each controller (or core) will execute one or  few iterations of the algorithm until the required approximation of the primal graph is achieved. 

There are several settings that we consider. The first possible setting is one controller that is responsible for the whole network with complete information about the network topology. The controller can perform the calculation on its own, possibly in parallel if it is multicore. In this case, there is no no need to exchange messages with other controllers and the information with regard the network topology is centralized. The other possible setting is that of several controllers where each is responsible for a particular sub-network, while they construct to the whole network. The construction itself is not distributed, i.e., each controller needs to know only the final construction. Instead of leader election, any participant suggests a random \textit{seed} and the actual seed used is the xor of the suggestions by other controllers. Each controller can reconstruct exactly the same overlay network.

The last setting that we consider is that of several controllers and the construction is distributed among all of them. In such case, each controller is responsible for a different random walker for constructing one tree at a time when instructed by the main controller. The random walker has a memory of size $n-1$. When a random spanning tree is generated, a message of size $n-1$ with the selected edges is sent to the main controller.

In such a setting, a monitoring algorithm for verification of expander construction is required. The monitoring algorithm can be executed by the main controller in charge of the network (see Algorithm~\ref{alg:alg4}). 

\begin{algorithm}
	\small
	\caption{\small Mixing Rate Based Monitoring}
	\label{alg:alg3}
	\DontPrintSemicolon
	
	\Input{$U_G^k$ for some $k$}
	\Output{$j$: number of counted nodes}
	\BlankLine
	
	$L$: max length of the walk \\
	
	$counter \leftarrow 0$ \\
	$v$: arbitrary chosen vertex \\
	$length \leftarrow 1$ \\
	$\Gamma'_{U_G^k} = \{\}$ \\
	\Repeat {all vertices are visited or $length > L$} { 
		\If {$v$ was not visited before} {	
			$counter \leftarrow counter+1$ \\
			$\Gamma'_{U_G^k} \leftarrow \Gamma'_{U_G^k} \cup \Gamma'(v)$ \\
		}	
		$length \leftarrow length + 1$ \\
		choose $u \in \Gamma'_{U_G^k}$ \\
		$v \leftarrow u$ \\
	}
	\Return $counter$ 
\end{algorithm}

We exploit the mixing rate based monitoring algorithm described in \cite{dolev_spanders:_2010} which is modified for our construction scheme (Algorithm \ref{alg:alg3}). It is known that computing the expansion of a graph is NP-hard. Thus, we utilize the fast mixing property, $\bigO(\log n)$, and the cover time, $\bigO(n \log n)$, of expander graphs in the monitoring algorithm.
The control software starts a random walk of length $\bigO(n \log n)$ on an arbitrary vertex $v$. Each new visited vertex by the random walk process is marked and counted, and its neighbors are added to a set of not yet visited vertices. Afterward, the walk proceeds from one of the randomly chosen neighbors. When the random walk is terminated, the counter is examined by the control software. In case the walk covered less than $n$ nodes, we can conclude with high probability that the graph is not rapidly mixing as was required or that there are too many edges in the constructed graph, implying that the construction was not successful \cite{dolev_spanders:_2010}.

Algorithm \ref{alg:alg3} can be sped up by performing parallel $\bigO(n)$ random walks of length $\bigO(\log n)$~\cite{dolev_spanders:_2010,alon_many_2008}. The controller software which runs the algorithm in SDN, can perform parallel random walks and possibly delegate to other controller units. Each of the controller units will return the calculated result to the main controller, upon whom will be the decision to stop generating random spanning trees.

Algorithm \ref{alg:alg4} is executed until the expansion of $U_G^k$, measured by the mixing rate quality, reaches the required value or at most $\bigO(\log n)$ iterations. The number of generated spanning trees is doubled after each iteration in order to approximate the expansion quicker and reduce the total number of iterations. 
%
\begin{algorithm}
	\small
	\caption{\small Distributed Expander Overlay Construction}
	\label{alg:alg4}
	\DontPrintSemicolon
	
	\Input{$G=(V,E, \omega)$}
	\Output{union of $k$ random spanning trees $U_{G}^{k}$}
	
	\BlankLine
	$U_{G}^{k}=\{\}$ \\
	$k \leftarrow 1$ \\
	\While{mixing rate requirement is not satisfied} { 
		$\mathcal{T} \leftarrow$ delegate the generation of $k$ random spanning trees \\
		\For {T in $\mathcal{T}$}
		{
			$U_{G}^{k} \leftarrow U_{G}^{k} \cup T$ \\
		}
		verify mixing rate requirement using Algorithm \ref{alg:alg3} \\
		$k \leftarrow 2 \cdot k$ \\
	}
	\Return $U_{G}^{k}$
\end{algorithm}

\section{Sparse Expander Graphs for Monitoring}
\label{sec:exp_graph_mons}
%

SDN allows for easy implementation of custom routing. We are interested in augmenting our overlay network with such custom routing, namely, a randomized routing. We show how such enhancement increases the level of anonymity or monitorability in the network. We describe two different random routing methods and assess the impact of each method on the desired network properties.

\subsection{Message Sending Scheme}
\label{sec:msg_scheme}
We utilize known properties of expander graphs: (a) expanders have many disjoint short paths \cite{kleinberg_short_1996}, (b) the diameter (specified by $diam(G)$) of an expander graph is $\bigO(\log n)$, and (c) expanders mix fast \cite{hoory_expander_2006}.

We propose a message sending scheme similar to the one proposed by Valiant in~\cite{valiant_scheme_1982}. Source node $s$, that sends a message to destination $d$, performs a random walk of length $\log n$ to an intermediate destination $v_1 \in V$. After arrival at $v_1$, another random walk of length $\log n$ is performed to an intermediate destination $v_2 \in V$. This is repeated for $r$ times, after which the message is sent directly to the destination $d$.
	%
	Each path between the vertices $\{s, v_1, \ldots, v_r\}$ is a random walk of length $\bigO(\log n)$. The last route from $v_r$ to $d$ is of length $\bigO(\log n)$ either since the diameter of the graph is $\bigO(\log n)$.
	Thus, overall route with $r$ intermediate nodes is of $\bigO(\log n)$ overlay edges, or, taking into account physical path, overall route complexity is $\bigO(\log ^2 n)$ edges. 

The constructed expander-based overlay network most likely be of an irregular degree. As a consequence, some nodes are more likely to be visited by a random walk process~\cite{newman_measure_2005}. We reduce such bias towards high degree nodes by a method similar to the one proposed in \cite{le_concentration_2017} (see also random walk with Metropolis filter in~\cite{lovasz_random_1993}). Roughly speaking, we can systematically assign a weight to each edge, so that a random walk process will behave as in a regular graph. As such, the set of vertices visited by a length $t$ random walk on an expander graph would be a randomly chosen sequence of $v_0, v_1, \dots, v_{t-1}$, where each $v_{i+1}$ is chosen uniformly at random and independently among the neighbors of $v_i$, for $i=0,\dots, t-1$~\cite{chung_spectral_1996}. The last route form $v_r$ to $d$ is a non-random walk based forwarding. However, the vertex $v_r$ is different and random at each path instance. Therefore, the randomized property of the overall message sending scheme is not impaired.



We suggest utilizing two routing methods: incremental path routing and loose routing. In the incremental path routing method, the sender's application proxy software builds the route hop by hop, corresponding and exchanging communication-related information (such as encryption keys) with each router on behalf of the user. As such, the sender's application proxy is aware of each participant in the chosen route. In the loose routing method \cite{goldschlag_hiding_1996}, any node along the message path may decide to alter the path, which means that the sender may be unaware of all the nodes which constitute the whole path. 

In case of loose routing, it is possible for the sender to route a message to a sub-area of the network where the local routing policy may decide upon a different routing method. The decision can be local to the sub-area or for the rest of the communication. Conversely, the implementation is more straightforward in the former case. As well, if encryption exists then the sender can negotiate session keys with each router it chooses on the route. As opposed to the latter case, it is left to the router that performs the modification to the route on behalf of the sender. 

Next, we evaluate the probability of successful monitoring in each routing method. For the following discussion, suppose that the set $C \subset V$ is a set of non-monitoring (compromised) nodes of size $\gamma < n$.

\subsection{Incremental Path Routing}
\label{sec:inremental_path}
In incremental path routing, it is assumed that the message path is a simple path. Hence, for each next node $v_i$ the probability for this node to be a non-monitoring node is $\frac{\gamma-i}{n-i}$. The probability for the overall route to be not monitored, i.e., composed of oblivious nodes only, is given by 

$$Pr[oblivious]=\prod_{i=0}^{r \cdot \log n}\left(\frac{\gamma-i}{n-i}\right)$$
Therefore, an overall route will have at least one monitoring node with probability
$$Pr[\overline{oblivious}]=1-Pr[oblivious]=1-\prod_{i=0}^{r \cdot \log n}\left(\frac{\gamma-i}{n-i}\right)=1-o(1)$$
The greater the route length, the faster the subtrahend in the last term goes to zero, and consequently, the probability of finding a monitoring node on a path increasing. 

%

\subsection{Loose Routing}
\label{sec:loose routing}

Let us assume that a walk of length $t$ begins at a node $v_0$ and advances to $v_1, v_2, \ldots,\allowbreak v_{t-1}$. Also, let us denote by $(C,t)$ the event that a random walk of size $t$ is confined to $C$, i.e., $v_i \in C$, $0 \leq i \leq t-1$. Suppose that $\gamma = \alpha n$, $\alpha < 1$. From~\cite{hoory_expander_2006} we have the probability for the event $(C,t)$, which is
$$Pr[(C,t)] \leq \alpha^t$$
This tells us that, it is exponentially unlikely for the random walk to traverse only non-monitoring nodes. 
We can estimate the size of the set $C$, assuming that we want the event $(C,t)$ to be at most $0.5$ for a given $t$:

\[
\left(\frac{\gamma}{n}\right)^t \leq 0.5 \rightarrow \gamma \leq \sqrt[\leftroot{-2}\uproot{2}t]{0.5} \cdot n
\] 
%

We can conclude from the above analysis that in a route of length $\bigO(\log n)$ at least one node is a monitoring node, and $\bigO(n / \log n)$ monitors are required in order to monitor the traffic in the network with high probability. Note that, the overall route length plays a significant role here. Increasing the route length reduces the required number of monitoring nodes and vice versa.

The above calculations are slightly relaxed. We can improve their accuracy using Chernoff bounds, which can be applied unaltered to hypergeometrically distributed binary variables~\cite{doerr_analyzing_2011}. Let us define an indicator variable $X_v$ for $v \in V$, which equals $1$ if a node is a monitoring node and $0$ otherwise:
%

\[
X_v = \left\{\begin{matrix}
1 & v \notin C \\ 
0 & otherwise
\end{matrix}\right.
\]
Let $p_i$ be the probability that $X_i=1$, and $r$ is the number of intermediate nodes. Let $X=\sum_{i=1}^{t}X_i$, where $t \in \bigO((r + 1) \cdot \log n)$ is the total length of a message path. It is easy to see that $\mu = E(X) = t \cdot \frac{n-\gamma}{n}$. Let $0 < \delta < t - \mu$, then 
\[
Pr[X \leq \mu - \delta] \leq e^{-\frac{2 \delta^2}{t}}
\]
which shows that the probability of a path being completely unmonitored is exponentially small.
%

\subsection{Measuring Monitoring Success}
\label{sec:measure_mons_sccs}

In this section, we quantify the monitoring level of the enhanced system. Remember that we enhance our system by systematically making the random walk traversals over the overlay network behave as in a regular degree graph. We show that in our enhanced network construction it is possible to monitor network traffic uniformly. First, we use a method based on Bayesian inference proposed in \cite{troncoso_bayesian_2009} to try and estimate the distribution of hidden states in the system. Such distribution can be used in order to infer events of interest in the system, for instance, a communication relationship between two nodes. Second, we use information theory metrics, that were proposed in \cite{diaz_towards_2003} and independently in \cite{serjantov_towards_2003}, for measuring the entropy of the whole system. The entropy measurement is useful for obtaining an estimate on the \textit{observability} level in the system. 

\noindent\textbf{Monitoring and Traffic Analysis.}
We want each monitoring node to audit as much traffic as possible. In \cite{troncoso_bayesian_2009}, it was shown that the hidden state of a system, which is the internal state of each node, is described by the path each message has taken. Therefore, sampling hidden states of a system is equivalent to sampling paths that messages in the system have taken. The goal is then to determine the distribution of message paths in the system in order to obtain a distribution of the hidden states. We show that sampling traffic in our constructed network is as sampling from a uniform distribution; which signifies that our constructed network is optimized for monitoring purposes. 

According to the message sending scheme, a user that is interested in sending a message can select $r$ intermediate nodes. The number of intermediate nodes affects the total path length, $\mathcal{L}_m$, of a message $m$, and this length is a multiple of $\log n$.  As mentioned above, the vertices from the set of traversed nodes, give a length $\mathcal{L}_m$, are uniformly sampled. Therefore, the probability for choosing a valid sequence of nodes, $\mathcal{V}_m$, is given by 


%
$$Pr[\mathcal{V}_m \mid \mathcal{L}_m] = \frac{r \cdot \log n}{n_{}}$$
%
Finally, the user selects uniformly at random the possible values of $r$. Since the last route is of length $\log n$ too and begins from a random node, then we can consider the fixed value $r$ to include the last route too. Thus, the probability for selecting a path $\mathcal{P}_m$ is 
\[
\begin{aligned}
Pr[\mathcal{P}_m] &= Pr[\mathcal{L}_m] \cdot Pr[\mathcal{V}_m \mid \mathcal{L}_m] 
= \frac{1}{r} \cdot \frac{r \cdot \log n}{n_{}} = \frac{\log n}{n_{}}
\end{aligned}
\]
and the probability of the hidden states in a system with $n_{msg}$ messages would be 
$$Pr[\mathcal{HS}] \propto \prod_{m=1}^{n_{msg}}Pr[P_m] = \prod_{m=1}^{n_{msg}}\frac{\log n}{n_{}}$$

Notably, the probability for every path of a message is equally likely, and consequently, the probability for each hidden state of the system would be equally likely too. Therefore, a user in our constructed network system cannot avoid monitoring by traffic analysis or by inner states distribution exploitation. Each monitor can uniformly audit the network traffic, and the network traffic is unbiased towards flowing through specific nodes.

\noindent\textbf{Entropy of the System.}
\label{sec:mon_avoid}
The maximum degree of monitoring is achieved when the sender sees all nodes as equally likely of being auditors of a message. The entropy of the system, after an external observer performs an observation, is compared against the maximum entropy~\cite{diaz_towards_2003}. This comparison gives a hint to the amount of information that was learned by the observer.

The results above show that the message paths are uniformly distributed and so are the hidden states of the system. Thus, we obtain that the entropy of the system is maximized, and each of the nodes is assigned an equal probability of being an auditing node. 



\noindent {\bf Monitors Locations.} 
The location of a monitor can be critical in a system in which most routing paths utilize a central vertex. Assuming that the underlying network does not have such central vertices by its nature (for example, a vertex connecting two sub-networks in a bridge network), in our proposed network construction no location for a monitor is more beneficial than the other. We estimate the monitoring potential of each node in the network by employing routing betweenness-centrality (RBC) measure which was proposed by Dolev et al.~in~\cite{dolev_routing_2010}. RBC of a vertex represents its potential to monitor and control data flow in a network. 

RBC can be defined when routing decisions depend on the target and are source-oblivious. For calculating the RBC value for each node, we inspect the number of packets that are expected at each node on the route to the destination. For a single node and a single message, the expectancy is dependent on the probability of a message arriving at a specific node. The message is sent along one of $\bigO(\log n)$ paths in the network. The probability of the message being routed to a particular node in the network graph is proportional to the probability of that node being part of the selected route. This probability is equivalent for all vertices in the constructed overlay network. As such, we obtain the result that the RBC is equivalent for every node in our constructed network; there is no preferred location for a monitor and each node has the equivalent potential of monitoring the network.

\section{Anonymity}
\label{sec:anonymity}

The anonymity of a subject is defined as being not identifiable within the anonymity set~\cite{pfitzmann_anonymity_2001}. The anonymity set is the set of all possible subjects who might cause an action. A subject is identifiable if we can get a hold of information that can be linked to the person. Similarly, as before, we use the entropy measurement of a system to estimate the information gained after an attack is performed on a system. The maximum degree of anonymity is achieved when an attacker identifies all users as equally likely of being the intended subjects of a message. 

The discussion in Section \ref{sec:mon_avoid} can be applied here in order to estimate the anonymity in the system. Similar reasoning leads to the result that all nodes in the network have an equal probability of appearing as relay nodes. Consequently, obtaining a maximum entropy, and as such, a maximum degree of anonymity.

Similarly to monitoring (Section \ref{sec:msg_scheme} and discussion in Sections \ref{sec:inremental_path} and \ref{sec:loose routing}), we can have more than half of the nodes compromised and yet remain anonymous in the network. 

\subsection{Anonymity On-Demand}
\label{sec:anonymity_on_demand}
In this section, we show how various features of our network architecture can mitigate common attacks on onion routing \cite{goldschlag_onion_1999} based anonymity networks. Also, we show how due to the flexibility of SDN the on-demand feature facilitates mitigation of certain attacks.

Intersection attacks are used to narrow down the possibilities of suspects among a monitored user-set by looking up the set of active users in the system. An intersection of these groups allows for the identification of communicating participants. Mitigation of these attacks can be achieved by using cover traffic; upon each output message, the node sends a corresponding fake packet on all of its incident edges. Since the constructed overlay network is of a bounded degree, $\bigO(\Delta n)$ dummy packets are required.


Predecessor timing attack assumes a set of relays that cooperate among themselves~\cite{wright_predecessor_2004}. Through introduction of delays in packets flow, attackers can identify whether they are on the same path, particularly, whether one of the attackers follows the victim on its path.

Unknown path length can significantly decrease the success of this attack \cite{wright_predecessor_2004}. In our message sending scheme, the number of intermediate nodes varies, possibly in combination with loose routing, this results in path lengths variation. Dynamic message paths decrease the likelihood for completing the attack or at least impair the confirmation of the attack's success. The amount of rounds required to perform the attack with high probability is $\bigO\left((\frac{n}{c})^2 \ln n\right)$ where $c$ is the number of attackers \cite{wright_predecessor_2004}. The required amount of rounds is greater than the route length, which is the order of $\bigO(\log n)$. As a result, until the attack is completed the anonymity network service will be shut down since messages have already arrived at their destinations.

Circuit clogging, or congestion attack, reveals a connection path by overloading a particular relay node and observing the affected circuits \cite{erdin_how_2015}. Those sorts of attacks, or other DoS attacks, can be regarded as if some nodes are compromised. We have seen in Section \ref{sec:mon_avoid} that more than half of the nodes in the network can be compromised.

The policy we propose prevents those attacks. The message sending scheme, using Valiant randomization technique, balances the load on the relays and links of the network. It is highly unlikely to congest a message path due to the randomization policy of the network. Additionally, loose routing can impair the success of the attack by dynamically modifying a path.

\section{Discussion}

We have presented a method for constructing a flexible, on-demand network service over SDN-enabled architecture. The network architecture has features of expander graphs and employs a random-routing policy for sending messages. We have shown that using such architecture allows us to monitor with high probability the entire network traffic, requiring a relatively small number of monitors. While the network architecture keeps being simple in construction, so that service providers, commercial companies, or private users can easily deploy it on-demand.


In particular, we have shown that $\bigO(n / \log n)$ monitors are enough to cover the entire overlay network when each path is of length $\bigO(\log^2 n)$ physical edges. The overlay network can be constructed in a distributed manner, converging faster towards the required features. The construction is achieved through random spanning trees generation. We have shown that the constructed overlay network graph is a sparse connected graph, which spectrally approximates the primal weighted graph if $\bigO(\log n / \epsilon^2)$ trees are included. Consequently, monitors located on the vertices of the overlay network can sample the substrate network traffic within $(1 \pm \epsilon)$ factor.

Further, we have shown that the network flows in our constructed architecture are uniformly dispersed resulting in a high level of monitorability. Namely, each monitor is equally likely to be exposed to the network traffic. Additionally, we have identified that applying our construction method results also in a high level of anonymity. 

Uniquely, this work presents the first of its kind, as far as the authors know, anonymity on-demand network service. The NaaS architecture of the network can mitigate most of the known attacks on anonymity networks. Notably, attacks that congest the network or result in relay server denial of service can be coped with the on-demand nature of the network. If the user, who deployed the anonymity network, suspects that over time the relays are compromised, they can periodically shut down the service and re-deploy it with a different network topology. This kind of service can be suitable for private communication use among trusted parties.

\begin{acks}
The research was partially supported by the Rita Altura Trust Chair in Computer Sciences; The Lynne and William Frankel Center for Computer Science; the grant of the Ministry of Science, Technology and Space, Israel, and the National Science Council (NSC) of Taiwan; the Ministry of Foreign Affairs, Italy; Infrastructure Research in the Field of Advanced Computing and Cyber Security; and the Israel National Cyber Bureau. 

We thank Noga Alon for the elaborate discussion and valuable comments. We are thankful to the anonymous reviewers whose comments helped improve this manuscript.
\end{acks}

\bibliographystyle{ACM-Reference-Format} 
\bibliography{network_lib}

\end{document}